\documentstyle[11pt,aaspp4]{article}

\def \cm3    {\mbox{$\rm cm^{-3}$}}
\def \H2     {$\rm H_{2} $ }
\def \be     {\begin{equation}}
\def \ee     {\end{equation}}
\def \rprime {r^{\prime}}
\def \la     {\langle}
\def \ra     {\rangle}
 
\begin{document}

\title{TIDALLY COMPRESSED GAS IN CENTERS OF EARLY TYPE AND ULTRALUMINOUS 
   GALAXIES} 
\author{Mousumi Das}
\affil{Raman Research Institute, C.V.Raman Avenue, Bangalore 560 080, India}
\centerline{email : mousumi@rri.ernet.in}

\centerline{and}

\author{ Chanda J.Jog}
\affil{Department of Physics, Indian Institute of Science, Bangalore 560012, 
India}
\centerline{email : cjjog@physics.iisc.ernet.in}

\begin{abstract}

In this paper we propose that the compressive tidal field in the centers of 
flat-core early type galaxies and ultraluminous
galaxies compresses molecular clouds producing dense gas obseved in the 
centers of these galaxies.  
 The effect of galactic tidal fields is usually considered disruptive in the 
 literature. However, for some galaxies, the mass profile flattens towards 
 the center and the resulting galactic tidal field is not
disruptive but instead it is compressive within the flat-core
 region. We have used the virial theorem to determine
the minimum density of a molecular cloud to be stable and gravitationally
 bound within the tidally compressive region of a galaxy. We have 
applied the mechanism to determine the mean molecular cloud
densities in the centers of a sample of  flat-core, early-type galaxies
and ultraluminous galaxies. 

For early-type galaxies with a core-type luminosity profile, the tidal field 
of the galaxy is compressive within half the core radius. 
We have calculated the
mean gas densities for molecular gas in a sample of early type galaxies which 
have already been detected in CO emission and we obtain mean densities of
$\rm\la n\ra\sim 10^{3}-10^{6}cm^{-3}$ within the central 100pc radius.  
We also use our model to calculate the molecular cloud densities in the inner few
hundred parsecs of a sample of ultraluminous galaxies. From the
observed rotation curves of 
these galaxies we show that they have a compressive core within their nuclear  
region. Our model predicts minimum molecular gas densities in the range
$\rm 10^{2}-10^{4}cm^{-3}$ in the nuclear gas disks;  the smaller
values are  applicable typically for galaxies with larger core
radii. The resulting density values agree well with the observed
range. Also, for large core radii, even fairly low density gas ($\rm\sim 10^{2}cm^{-3}$)
can remain bound and stable close to the galactic center.

\end{abstract}


\noindent Running title: Tidally compressed gas in centres of galaxies

\section{INTRODUCTION}
 
Over the last two decades observations have revealed the presence of significant 
amounts of cold, neutral gas in early type galaxies (Henkel \& Wiklind 1997; 
Knapp 1998). The term "early type galaxies" includes 
lenticular or SO galaxies, dwarf elliptical (dE) and elliptical (E) galaxies. Many
of these galaxies have prominent dust lanes and extended distributions of neutral
hydrogen gas (HI). A significant fraction of these galaxies also contain molecular 
hydrogen gas ($\rm H_{2}$). Unlike the HI gas which is usually distributed in the  
outer parts of the galaxy, the molecular gas is nearly always centrally concentrated.
The \H2 gas in these galaxies has been detected and traced using the CO(1-0) 
and CO(2-1) lines, indicating that the gas may be fairly dense
($\rm\sim 10^3 cm^{-3}$). 
In a few cases high density molecular gas n$(\rm H_{2})>10^{4}cm^{-3}$ 
has been observed using the high density tracers HCN and HCO$^{\rm +}$.
The amount of molecular gas observed in early type galaxies  
varies between $10^{5}\rm M_{\odot}$ to $10^{9}\rm M_{\odot}$ and is less
than in spirals of the same blue luminosity (Knapp 1998).  While
the amount of cold gas is small compared to that in spiral galaxies,
it is still not insignificant and in some cases also forms the site of
ongoing star formation. The origin and spatial
distribution of this dense gas has not been explained so far.

Ultraluminous galaxies are very bright starburst galaxies having
far-IR luminosities
of the order $\rm L > 10^{12}L_{\odot}$ (Sanders et al. 1988). They are 
formed from the  merger of two gas 
rich spiral galaxies. The merger is expected to result in large amounts of
disk molecular gas falling into
the center of the newly formed merger remnant, which
subsequently results in enhanced star-formation  or
super-starburst at the center. 
The total central gas mass
is $\rm\sim 10^{9} - 10^{10} M_{\odot}$, and is distributed over the central
few hundred pc.    
 
It may seem strange at first that we are studying seemingly
different kinds of galaxies (early type galaxies and ultraluminous galaxies)
together in this paper. However, as we argue next, these are close in
terms of their dynamics. Mergers of two spirals are known to result in an
ultraluminous galaxy, as seen from observations (Sanders et al.
1988), and as explained theoretically (e.g., Jog \& Das 1992).
A merging pair is believed to be a precursor of an elliptical
galaxy, as shown observationally (e.g., Schweizer 1999), and in
fact the K-band brightness profiles of about 50 \% of the 
ultraluminous galaxies display a $r^{1/4}$ distribution typical of an 
elliptical galaxy (Sanders \& Mirabel 1996).
Numerical simulations have shown that a merger of two spirals 
can result result in an elliptical galaxy (e.g., Barnes \& Hernquist 1992). 
This scenario is supported by the
fact that the observed gas surface density in the centers of
ultraluminous galaxies is $\rm 500 M_{\odot} pc^{-3} $ and is comparable
with the central stellar density in ellipticals (Kormendy \&
Sanders 1992; Solomon 1997).  
Therefore, we expect that the mass distribution and hence the dynamics of 
the centres of these galaxies would be similar. Also,  
the gas number densitites would be 
affected in a similar way by the galactic tidal field. 
Hence we study the origin of the
dense gas in the centres of early type and ultraluminous galaxies together
in this paper. We do not, however, study the relevance of the
compressive tidal field, which we discuss next, for the evolution from 
a merging pair of spirals into an elliptical galaxy. 
 
The  galactic
tidal field is usually thought to be disruptive in the centers of galaxies 
so that only dense gas would
survive the strong shearing due to a central tidal field. 
However, if the potential flattens instead of peaking towards the center of 
the galaxy, the tidal field can reverse sign and become
compressive. 
The tidal field is compressive when the background gravitational
field of the galactic disk enhances the self-gravitational field
of the cloud (Binney \& Tremaine 1987, Chap. 7). We show this situation 
is physically possible in core regions of some galaxies - such
as the flat-core early-type and ultraluminous galaxies.
This is a new application of the idea of compressive tidal fields,
namely, in the astrophysical context of galaxy centers.
In the past, the effect of the compressive nature of tidal 
fields has been studied  
in the context of disk shocking of globular clusters (Ostriker, Spitzer,
\& Chevalier 1972), and the tidal effect on a galaxy in the core
of a cluster of galaxies (Valluri 1993). The other new feature
of our study is the application of compressive fields to a
compressible fluid, namely gas, and our calculation of the
steady-state density of a bound, virialized cloud in the presence of 
this field.

In \S 2 and \S 3,  we  determine the minimum density of a stable
cloud in a compressive 
tidal field. We have applied these results (\S 4) to the centers 
of a few flat-core, early-type  
galaxies in which molecular gas has been detected and also to a sample
of ultraluminous galaxies. We have compared our
results with the observed gas densities.  We thus show
that it is the compressive nature of the galactic tidal fields
in the centres of these galaxies that naturally leads to the
origin of the dense molecular gas observed in these galaxies.

\section{EQUILIBRIUM OF A CLOUD IN THE TIDAL FIELD OF A GALAXY}

Molecular clouds are generally bound by self-gravity, hence the compressive
 galactic tidal 
field plays an important role in the stability of a cloud. In this 
section we use the galactic tidal field in a spherical potential to
derive the net stability criterion for a cloud. 

\subsection{Tidal Field}

The components of the tidal field in a spherical potential has been 
derived in some detail by Cowsik \& Ghosh (1987), and Valluri (1993) who 
studied the effect of the 
compressive tidal field of a cluster potential on a galaxy.  We
summarize their results
 briefly here before applying them to a molecular cloud.
For a potential $\rm\Phi(X_{i})$ the tidal field components at $\rm x_{i}$, where
$\rm x_{i}=X_{i}-X_{io}$, are given by, 
\be
{\bf T_{x_{j}}} = -\left[\Sigma_{i}\frac{\partial^{2}\Phi}{\partial X_{i}\partial X_{j}}
\right]_{o}x_{i}{\bf\hat{x}_{j}}
\ee
\noindent For a spherical potential $\rm\Phi(R)$ for which the cloud lies along the 
X axis so that X=R, Y=Z=0, the components of the tidal field at a position (x,y,z)
in the coordinate system centered on the cloud, are given by,
\begin{eqnarray}
T_{x} = -\left[\frac{\partial^{2}\Phi}{\partial R^{2}}\right]_{o}x\\
T_{y} = -\frac{1}{R}\left[\frac{\partial\Phi}{\partial R}\right]_{o}y\\
T_{z} = -\frac{1}{R}\left[\frac{\partial\Phi}{\partial R}\right]_{o}z
\end{eqnarray}
\noindent Let $\la\rho\ra$ be the mean density of the mass within radius R in the 
galaxy so that $\la\rho\ra=\frac{3M(R)}{4\pi{R^3}}$, where M(R) is the mass of the galaxy 
within radius R. Then the tidal field components become,
\begin{eqnarray}
T_{x} = \frac{4}{3}\pi G(2\la\rho\ra - 3\rho)x\\ 
T_{y} = -\frac{4}{3}\pi G\la\rho\ra y\\
T_{z} = -\frac{4}{3}\pi G\la\rho\ra z 
\end{eqnarray}
\noindent Thus, the tidal field is always compressive along the directions perpendicular
to the radius vector for a spherical potential. It can be either compressive or disruptive 
along the radius vector depending on the density distribution in the galaxy. For 
$\rho(R) > \frac{2}{3}\la\rho\ra$ the tidal field is compressive.
The nature of the tidal field thus depends on the density profile of the system. For a
centrally peaked density profile, the tidal field will always be disruptive along the radial
direction. However, if the density profile of the galaxy flattens towards the center, then 
the tidal field can be compressive within the flat core region of the galaxy. The tidal 
field is compressive in the centers of clusters of galaxies and is important in the
evolution of galaxies in the cluster (Valluri 1993). We show that it is also
compressive in the centers of flat core early type galaxies and some ultraluminous 
galaxies; this is derived using equation (2) and is discussed in Sections 3 and 4 of 
this paper. Equations (5)-(7) are shown to illustrate the physics of compressive
tidal fields. For these galaxies the tidal field
changes direction from being a disruptive force to a 
compressive force in the center of the galaxy. We show that this is an important factor in 
deciding the steady state density of a cloud in the galaxy center.
 
When the tidal field is compressive along the line of action of the force, the cloud will 
feel an overall compressive force. In such cases the tidal field can never disrupt the cloud, 
instead the cloud will be compressed by the galactic field. For a cloud in equilibrium in
a compressive tidal field, the internal pressure maintains equilibrium against both the  
self gravity of the cloud and the compressive tidal field of the galaxy.

\subsection{Virial Equilibrium of a Cloud}

For a system in virial equilibrium, we have the relation
\be
-2\la K\ra = \sum_{i=1}^{n} {\bf{F_{i}}.{r_{i}}}
\ee
\noindent where $\rm\la K\ra$ is the time averaged kinetic energy of 
the system, $\rm\bf{F_{i}}$ is the force on the i$^{\rm th}$ particle  and $\rm{\bf r_{i}}$ 
is the position coordinate of the i$^{\rm th}$ particle wrt the cloud center. For a 
molecular cloud in virial equilibrium, the forces involved will
be the internal self gravity of the cloud and the galactic tidal field acting on 
the cloud. As mentioned previously, the tidal field is anisotropic. However for 
simplicity we will consider only the radial component and not the transverse 
components of the tidal field in the deriving the equilibrium density of a cloud. Strictly
speaking we should include all three components of the tidal field which would lead to 
an ellipsoidal cloud. However, in a real cloud, the compression
 would tend to become isotropic soon (see Section 5). Hence, for
simplicity, we assume that the cloud is spherical and we include only the 
radial tidal field for estimating the equilibrium cloud density.
If we convert the right hand side of the equation (8) into an integration, we obtain the 
following relation,
\be
-M_{c}v^{2} = -\int\frac{GM_{c}(r)dM_{c}(r)}{r} + T_{0}\int r^{2}dM_{c}
\ee
\noindent where $\rm M_{c}$ is the mass of the cloud, v is the 3-D velocity dispersion 
in the cloud and $\rm T_{0}$ is the tidal field per unit
distance across the cloud (i.e. $\rm T_{0} = -\frac{\partial^{2}\Phi}{\partial r^{2}}$).
We assume that the density within the cloud is inversely proportional to 
radius as observed (Sanders, Scoville, \& Solomon 1985), so that
$\rm\rho_{c}(r) = \frac{\rho_{o}R_{c}}{r}$, where $\rm\rho_{o}$ is the density of the cloud at 
the cloud edge, i.e. at $\rm r=R_c$. Then the cloud has a mass profile 
given by $\rm M_{c}(r) = 2\pi R_{c}\rho_{o}r^{2}$. Thus, the
mean density of the cloud,
 $\rm\la\rho_{c}\ra$, is obtained to be 
$\rm\la \rho_{c}\ra=\frac{3}{2}\rho_{o}$. The equation of cloud equilibrium thus 
reduces to,
\be
v^{2} = \frac{2}{3}\frac{GM_{c}}{R_{c}} - \frac{1}{2}T_{0}{R_{c}}^{2}
\ee

This has been derived in the galactic rest frame. In the
Appendix A, we give the derivation in the rest frame of the
cloud for the sake of completeness,
and argue that for considering cloud stability it is appropriate
to use the galactic frame equation as derived here.

\subsection{Minimum Density for Stability} 

For a tidally disruptive field, the tidal field coefficient $\rm T_{0}$ is positive. 
Since $\rm v^{2}>0$, we obtain the familiar relation for tidal stability of a molecular 
cloud, namely
\be
\frac{2}{3}\frac{GM_{c}}{R_{c}}>\frac{1}{2}T_{0}{R_{c}}^{2}
\ee
Writing this condition in terms of $\rm\la \rho_{c}\ra$, the
mean density of the cloud, we get
\be
\la\rho_{c}\ra > \frac{9}{16\pi G}T_{0}
\ee
Thus, the minimum density for a cloud to be stable against the 
 disruptive tidal field is 
\be
\la\rho_{m}\ra=\frac{9}{16\pi G}T_{0} 
\ee
For a tidally compressive field 
$\rm T_{0} = -\mid T_{0}\mid < 0$, hence $\rm v^{2}>0$ does not give us any new information
except that the cloud can never be disrupted by the tidal field. So to obtain the density 
of a cloud in the compressive tidal field we impose the
condition that the cloud be gravitationally
bound to begin with, i.e.
\be
\frac{1}{2}v^{2} - \frac{2}{3}\frac{GM_{c}}{R_{c}} \leq 0 \\
\ee
Using equation (10) we get 
\be
\la\rho_{m}\ra=\frac{9}{16\pi G}\mid T_{0}\mid
\ee
\noindent which is formally the same relation as that for a disruptive tidal 
field the only difference being that we take   
the magnitude of the tidal force coefficient. Physically, however they 
have very different significance. For a disruptive field, the minimun density
$\rm\la\rho_{m}\ra$ is the density
below which a cloud will be disrupted by the galactic tidal field. 
For a compressive tidal field 
it denotes the minimum density for a gravitationally bound cloud 
to exist within that
region.  Since the cloud is assumed to be gravitationally
bound to begin with, an additional inward/compressive field
increases the velocity dispersion required to satisfy the virial
equilibrium (eq.[10]), and hence the critical density required
for stability in presence of a compressive force is actullay
{\it higher} for a virialized cloud.
For lower densities, the gas within that region will be diffuse and 
gravitationally unbound. Such gas will not lead to star formation.

Note that the critical number density of hydrogen molecules (in $\rm cm^{-3}$) 
corresponding to the mass density  $\rm\la\rho_{m}\ra$ (eq.[15]), 
assuming a 10 \% gas number fraction in helium, is given by $\rm\la\rho_{m}\ra 
 / (4.68 \times 10^{-24}) $. We will use this conversion in the
next two sections. To calculate and plot tidal fields, we use mass, length
and time units to be $\rm M_{\odot}$, pc and $\rm 10^{7}$year respectively.
This means that the units of $\rm T_{0}$ are 
$\rm\left(\frac{1}{10^{7}year}\right)^{2}$. 

\section{APPLICATION TO MOLECULAR GAS IN EARLY TYPE GALAXIES}

In this section we apply the above criterion to predict the central molecular gas 
densities in six early type galaxies. As mentioned later in this section, we need 
parameters that can be derived from the luminosity profile. Unfortunately, only a few 
of the galaxies whose surface brightness has been studied have also been detected in 
CO. Hence we could apply the results of the previous section to only a small sample of 
early type galaxies. 

\subsection{Spherical Stellar Potential}

Tremaine et al. (1994) have derived a one parameter set of models for spherical 
stellar systems, these give self-consistent potential-density pairs. Their 
model can be applied to the center of early type galaxies and 
spiral bulges. From their model they derive the density distribution, surface 
brightness profile and line of sight velocity dispersion. For all the models, at large 
distances the density distribution tends to $\rm \rho(r) \propto r^{-4}$ and the surface
brightness tends to the familiar $\rm r^{1/4}$ form. The family of potentials, called 
$\rm \eta $ models are given by,
\be 
\Phi(\rprime) = \frac{GM_g}{R_{o}}\frac{1}{(\eta-1)}\left[\frac{{\rprime}^{(\eta-1)}}
{(1+{\rprime})^{\eta-1}} - 1\right]
\ee
\noindent where $\rm\rprime=R/R_{o}$, $\rm 0<\eta\leq 3$, 
$\rm M_g$ is the mass of the galaxy and $\rm R_0$ is the core radius. Many of the 
properties of the 
$\rm\eta $ models have also been derived by Dehnen (1993). 
The $\rm\eta=1$ model corresponds to the Jaffe (1983) model while the $\rm\eta=2$ model 
corresponds to the Hernquist (1990) model. From equation (16), $\rm T_0$,
 the tidal field per unit distance is found to be,
\be
T_{0} = -\frac{GM_g}{R_o}\frac{{\rprime}^{(\eta-3)}}{(1+{\rprime})^{\eta+1}}(\eta-2-2\rprime)
\ee  
For $\rm 0<\eta\leq 2$ the density diverges towards the center of the galaxy and the tidal 
field is always disruptive. In contrast,  for  $\rm 2<\eta\leq 3$ the model 
has a more flat density core
and the tidal field is compressive within the core. However,
even here, the field is compressive only for distances 
$\rm R<\frac{1}{2}R_{o}$. 

Thus to determine the critical density in the centers of these galaxies we need to 
determine two parameters; one is $\rm M_g$, the mass of the galaxy, which can be determined 
from the mass to luminosity ratio (M/$\rm L_{v}$) and the  
other is the observed core or break radius $\rm R_{o}$. The break
radius can be determined from the surface brightness profile of
the galaxy. For a flat core 
profile the break radius is very sharp but for a power law profile, which is rising 
towards the center, $\rm R_{o}$ is not so sharply defined.

\subsection{Results for Minimum Gas Densities} 

There are now a very large number of early type galaxies with detected \H2 gas 
(Knapp 1998; Henkel \& Wiklind, 1997). In fact, after looking at the literature,
we found that there are over fifty elliptical and SO galaxies with
detected \H2 gas. Many show both CO(1-0) and CO(2-1) lines towards 
the center indicating that there may be fairly dense gas in these galaxies.
The fact that dense gas does exist in many early type galaxies
indicates that there must be some stabilising mechanism resulting in fairly long 
lived dense gas in the
centers of these galaxies. 

In this section we use the results of the previous section to
explain the observed molecular gas densities in a sample of elliptical and SO galaxies. 
To determine $\la\rho_{m}\ra$ we need both the galaxy mass and the break radius to apply
the analytical models of Tremaine et al. (1994). These parameters have been determined
for a large sample of early type galaxies by Faber et al. (1997). They used very high 
resolution HST observations to determine the central parameters of 61 early type galaxies.
We compared their sample with CO observations of elliptical galaxies in the literature. We
found that molecular gas has been detected in the center of only a few of the galaxies 
in the Faber et al. (1997) sample.   
We have used the data in column 3 and 5 of Table 3 in Faber et al. (1997) to determine 
the mass of the galaxy $\rm M_g$ and we used the data in Table 2 of their paper to 
determine the break or core radius $\rm R_{o}$ of the galaxy. The results are shown in Table 1 
where we have listed the galaxies, the parameters $\rm R_{o}$ and $\rm M_g$ and the 
density of molecular gas predicted from equation (15) of the previous section. 
Figure 1 illustrates how the tidal field varies in a power law galaxy (NGC 4697) and
in a core profile galaxy (NGC 4649). For the latter, the tidal field becomes negative
i.e. compressive within a certain radius in the galaxy, while for NGC 4649 the tidal
field continues to rise towards the center. Figure 2 shows
the variation of miminum density in the disruptive and compressive tidal regions of
NGC 4472. The dashed line demarcates the regions of the compressive and disruptive 
tidal field.  
In the following paragraphs, we have compared our results with the CO observations of the
sample of galaxies taken from Faber et al. (1997).

\noindent NGC 4594 and NGC 4697 : Both these galaxies have power law profiles and thus the
tidal field is always disruptive in both of them. NGC 4594 (also called `Sombrero Galaxy') 
has a very steep brightness profile in the center and also has an active nucleus 
(Bajaja et al., 1988). Bajaja et al. (1991) observed this galaxy in the CO (1-0) 
and (2-1) line and though they detected CO in the disk of the  galaxy, they could not detect 
any CO in the center. The reason for this could be that the tidal field is very high in the 
inner few hundred pc and the gas cannot survive there. NGC 4697 on the other hand has a 
moderately steep luminosity profile with a break radius of $\rm\sim 240pc$ 
(Faber et al. 1997). Using $\rm\eta=1.5$ in equation (17) we get the tidally limited density 
to be $\rm\la n\ra >\sim5\times10^{2}cm^{-3}$ at 1 kpc and it rises to 
$\rm\la n\ra\sim10^{5}cm^{-3}$ at 200pc. 
Sofue and Wakamatsu (1993) have detected CO emission from the central 1 kpc of this galaxy.
Since the  mass profile does not rise very steeply in the center of this galaxy, the tidal
field is not very strong and so gas can exist in the center of NGC 4697 even though it has 
a power law profile like NGC 4594. We do not address the origin
of high density gas observed in some power-law galaxies.

\noindent NGC 4472 : This galaxy has a flat core luminosity profile and hence the 
tidal field is compressive in the center. The break radius is $\rm\sim$180pc which 
means that gas will feel the compressive tidal field within the inner 100pc where 
densities of at least $\rm\la n\ra\sim10^{4}cm^{-3}$ are predicted by equation (15). CO emission
detection indicates that there is $\rm \sim10^{7}M_{\odot}$ of molecular gas in this 
galaxy (Huchtmeier et al. 1994; Lees et al. 1991). However, Huchtmeier et al. (1988)
could only detect CO from an off center position and not at the center. One 
reason for this could be that their beam was very large ($\rm\sim 2kpc$). Higher
resolution observations might lead to detection of high density gas in the center of 
this galaxy.  

\noindent NGC 4649 : This galaxy is a giant elliptical belonging to the Virgo cluster. 
It has a fairly large core which is compressive at radii less than 130pc. The model predicts gas
densities of at least $\rm <n>\sim5\times10^{3}cm^{-3}$ in the inner 130pc. CO(2-1)
and CO(1-0) emission has been detected from this galaxy (Lees et al. 1991; Sage \& Wrobel,
1989). 

\noindent NGC 1400 \& NGC 1316 : These galaxies both have small cores and the model
predicts stable gas of densities of at least $\rm\la n\ra>\sim 10^{5}cm^{-3}$ to $\rm 10^{6}cm^{-3}$ 
in the central 10 to 20pc of these galaxies. If the galaxy had a power law profile the
gas would probably not be stable so close to the center. CO emission has been detected 
from NGC 1400 by Lees et al. (1991) and in NGC 1316 by Sage and Galletta (1993).

\noindent NGC 1600, NGC 3379 \& NGC 4486 : These  galaxies have not been 
detected in CO, only upper limits to their molecular gas masses have been obtained 
(Sofue \& Wakamatsu, 1993). All three galaxies are ellipticals with  compressive 
cores and hence dense gas can exist in the inner 100pc of their cores.

\section{APPLICATION TO MOLECULAR GAS IN ULTRALUMINOUS GALAXIES}

Ultraluminous galaxies are merger remnants with large amounts of molecular gas 
concentrated in their centers. Over the past few years observations have 
revealed that the nuclear molecular gas is located in a central disk of size  
smaller than a few hundred pc (Scoville et al. 1991; Downes, Solomon 
\& Radford, 1993). Downes \& Solomon (1998) observed with high resolution 
a sample of ultraluminous galaxies and showed that most of the CO flux from 
the center comes from subthermally
excited CO emission. They concluded that most of the gas is of density 
$\rm\sim10^{3}cm^{-3}$ and only $\rm\sim$25\% of it is high density gas 
($\rm\geq 10^{4}cm^{-3}$).  We  use our model to determine the minimum density 
of bound, virialized molecular gas clouds for
their sample of galaxies using their parameters for the
rotation curves, and show that our resulting values compare well
with their observations. Despite the high observed turbulent velocities (see Table 5,
Downes \& Solomon, 1998), the assumption of an individual cloud being bound may 
still be applicable if the beam covers a number of clouds and hence a large fraction 
of the linewidth is due to cloud-cloud velocity dispersion. Also, the volume filling factor 
of the central molecular gas is $\sim$ 1. Nevertheless, we have treated the central 
gas as being discrete bound clouds for simplicity. 

\subsection{Tidal Field from the Nuclear Rotation Curve}

Downes \& Solomon (1998) have observed the rotation curves of the molecular gas 
disks in a sample of ten ultraluminous galaxies. They found that the rotation curves 
rose to a steady velocity $\rm V_{o}$ at a radius $\rm R_{d}$ and then became flat 
at the steady value. They modelled the rotation curves with the simple 
expression,
\be
V_{r}(R)=V_{o}\left(\frac{R}{R_{d}}\right)^{\beta} 
\ee
\noindent where $\rm\beta=1$ for $\rm R<R_{d}$ and $\rm\beta=0$ for 
$\rm R\geq R_{d}$ upto the maximum radius at which CO was observed.

For the power law rotation curve (for $R<R_{d} $), the corresponding potential
can be calculated to be:
\be
\psi (R) = \frac {{V_d}^2}{2\beta}(\frac{R}{R_d})^{2\beta}
\ee                             
Hence, $\rm T_0$, the tidal field along the radial direction is obtained to be:
\be
T_{0}=-(2\beta-1)\frac{{V_d}^2}{{R_d}^2}
\ee  
Note that $\rm T_{0}<0 $, that is the radial component of the tidal
field is compressive, when $\rm\beta>{1/2} $. Therefore, for
$\rm\beta =1$ as seen in the inner regions of ultraluminous
galaxies, the galactic tidal field is compressive.  
For the region outside of $\rm R_d$ where the rotation curve is
flat, the galactic tidal field is not compressive, instead it is
disruptive.

In Figure 3, we show the plot of the tidal field per unit mass
versus the galactic radial distance for Arp 193 for the data
from Downes \& Solomon (1998). The tidal field
is compressive inside of 220 pc. Note that 
sharp transition from disruptive to compressive region is an artifact 
due to the rotation curve model obtained by Downes \& Solomon
(1998), the actual transition is probably smoother.

\subsection{Results for Minimum Gas Densities}

We used the equation (15) and equation (20) to
determine the minimum density for a bound, virialized cloud to exist in the 
compressive tidal field of the ultraluminous galaxies observed by Downes \& 
Solomon (1998). Our results are shown in Table 2 where we have listed the 
galaxies, their rotation curve turnover velocity, turnover radius and the 
minimum gas density $\rm<\rho_{m}>$.
The particle number density $\rm\la n\ra$ lies in the range of $\rm 10^{2}$ to 
$\rm 10^{4} cm^{-3}$. This density is the {\underline{minimum}} density of a 
gravitationally bound cloud to remain stable in the center of the galaxy. 
This range agrees very well with the mean density of the main mass component 
of the molecular gas observed in these galaxies (see Table 6, 
Downes \& Solomon 1998). Since we have assumed discrete clouds with a volume 
filling factor $<$ 1 (\S 4), our values are slightly larger than the observed
values, as expected, since the latter have been derived assuming
a uniformly filled disk. 

The peak density of the molecular gas is observed to be near
the turnover point in the rotation curve whereas our model
predicts a constant high density inside this point.
Also there is substantial gas observed
beyond the turnover point, but both these could arise because of
the sharp turnover in the model rotation curve assumed by 
Downes \& Solomon (1998).

Note that the very dense gas with $\rm n \geq 10^{4} cm^{-3} $
which constitutes about $25 \%$ of the gas mass in the extreme
starburst regions such as Arp 220 West and East could also arise
due to tidal compression (see Table 12, Downes \& Solomon 1998
and our Table 2). This high density gas  is
the site of HCN emission, and contains a large fraction of the star formation.
Since the internal velocity dispersion increases with density for a bound
virialized cloud, the 
observed velocity dispersion in a given region sets an upper
limit on the high density that can result from this mechanism.
The observed dispersion in a region could be due to the
superposition of a number of moderate density clouds
($\rm\la n\ra\sim 10^{2}-10^{4}cm^{-3}$) in the beam, as we argued in \S 4,
or it could be due to a few higher density clouds. We do not discuss
the issues of the fraction of the gas in the dense component,
and the origin of the high central turbulent velocities, in this paper.

It is interesting that even low to moderate 
density gas $\rm\la n\ra\sim 10^{2}-10^{3} cm^{-3}$  in some cases
can survive in the center and will not be disrupted by the galactic
tidal shear. Downes \& Solomon (1998) suggest that most of the  moderate 
density gas ($\rm 10^3 cm^{-3}$) that they observe in these galaxies is diffuse and unbound as it 
may not be able to withstand the tidal shear of the nuclear region  of the galaxy. However as we
have shown here, gas can survive near the center as bound clouds because the tidal
field is compressive and not disruptive. Hence the gas that they have
termed as the `diffuse medium' could well be bound clouds.

Since these clouds are bound, they are more susceptible to the onset
of star formation. Thus, our
conclusion that these clouds are bound helps the
scenario proposed by Downes \& Solomon (1998) for starbursts 
powering the ultraluminous galaxies. The starbursts could be triggered
due to large-scale gravitational star-gas instability (Jog
1996), as they suggest, and/or it could arise due to the
collapse of the individual, bound clouds, or due to collisions
between the bound clouds.   Hence the formation of dense 
gas ($\rm\la n\ra\geq 10^{2}-10^{4}cm^{-3}$) in a compressive tidal field is 
important for the starburst and hence the evolution of ultraluminous galaxies. 

\section{DISCUSSION}

\begin{itemize}
\item[1.]  Normal spirals also show dense gas in their central regions
(e.g., Henkel, Baan, \& Mauersberger 1991),
so it is natural to ask if one could apply the above analysis and
obtain the resulting steady-state density of the dense, virialized clouds.
Carollo and Stiavelli (1998) have observed using HST the nuclear surface 
brightness profile of a large sample of spiral galaxies. A considerable 
fraction of these spirals have a flat core luminosity profile
close to their center. The galactic tidal field for these galaxies can
become compressive within the core. We plan to study these galaxies
in further detail in a forthcoming paper. In particular we would like 
to examine how the molecular cloud densities in these galaxies compare 
with those 
predicted by our work. This may help us decide whether the
so-called diffuse, inter-cloud medium in normal and starburst galaxies
(e.g., Bally et al. 1988; Jog \& Das 1992) actually consists of 
bound, virialized clouds. 
\item[2.] However, the mass distribution and hence the dynamics in the centres
of spirals is more complex due to the common presence of bars and triaxial
bulges. We find that for the Milky Way Galaxy, assuming the bar potential 
as in Binney et al. (1991), the field is not compressive. A typical bulge
potential (e.g., Sellwood \& Wilkinson 1993) 
also does not result in compressive fields. Physically, this
could be because a bar is a strong deviation from an
axisymmetric potential, and the field near the bar is not
smooth. In fact, for a molecular cloud close to a strong bar, the galactic
tidal field is shown to cause an internal heating of the cloud
(Das \& Jog 1995). We will discuss the barred case in a future paper.
\item[3.] We have only considered the radial tidal field in this paper.
The tidal fields along the other two orthogonal directions are
always compressive (see Section 2.1), and have magnitudes
comparable to the radial case. We are justified in neglecting this
asymmetry since the gas clouds are compressible and would tend to
have an isotropic velocity dispersion, and hence the net effect of
the compressive fields would be seen in an average sense.
In the case of the ultraluminous galaxies, this asymmetry would
be particularly stronger since the height of the disk
is small and hence using a 
3-D distribution is not strictly correct.
\item[4.] In the centers of ultraluminous galaxies, the gas constitutes a 
significant fraction of the mass ($\geq 25 \% $) and hence the gravitational
potential. Hence a gas cloud does not respond to the stellar
potential alone.  Since we have used the observed gas rotation
curve in obtaining the potential, we have used
the potential as seen by a gas cloud. 

\end{itemize}

\section{CONCLUSIONS}

We have investigated a simple but novel physical process which can compress 
gravitationally bound, virialized molecular clouds in the centers of galaxies.
The main conclusions of this study are the following.

\begin{itemize}
\item[1.] The galactic tidal field changes from being disruptive to 
compressive in the centers of flat core early type galaxies and 
ultraluminous galaxies. 
\item[2.] The compressive tidal force and the self gravity of the cloud
are balanced by the internal pressure of the cloud. We have used the
virial theorem to derive a simple expression for the minimum mean density
of a gravitationally bound cloud in the compressive core of a galaxy. 
The additional inward/compressive field
increases the velocity dispersion required to satisfy the virial
equilibrium, and hence the critical density required
for stability in presence of a compressive force is 
{\it higher} for a gravitationally bound, virialized cloud.
 For lower densities, the gas within that region will be diffuse and 
 gravitationally unbound. Such gas will not lead to star formation.
\item[3.] We have applied our results to a sample of flat core, early
type galaxies taken from Faber et al. (1997).  The gas densities 
predicted by the model varies with core size and cloud position in the
center of the galaxy, and the typical density range observed within 
the central 100pc is $\rm\la n\ra\sim 10^{3}-10^{6}cm^{-3}$. 
The galaxies in our sample were chosen so that
they have an observational evidence of molecular gas in their centers.
However, the spatial resolution of these data are not yet
adequate and only give a beam-averaged lower limit on the gas
densities. Future high resolution studies of such galaxies are highly
recommended.
\item[4] We have also applied our results to the molecular gas disks  
observed in the centers of ultraluminous galaxies (Downes \& Solomon, 1998).
From their observed rotation curves, we show that they have a compressive
core within their central region.
The minimum mean molecular cloud densities predicted from our model are 
in the range of $\rm 10^{2} cm^{-3}$ to $\rm 10^{4}cm^{-3}$ and
this range matches well with the 
observed gas densities in these galaxies. Also, if the tidal field is
compressive and not disruptive as previously thought in the literature,
this molecular gas can exist in the form of bound clouds. 
Hence, these clouds are closer to 
star formation than unbound, diffuse molecular gas. This strengthens the
scenario proposed by Downes \& Solomon (1998) for starbursts 
powering the ultraluminous galaxies.
\end{itemize}
 
\acknowledgments

M.Das would like to thank R.Nityananda for useful discussions 
during the course of this work, and we would like to thank M.
Valluri and S. Faber for very useful e-mail correspondence 
regarding this work. We thank the anonymous referee for helpful comments
which improved the comparison with observations of 
ultraluminous galaxies. We thank S. Sridhar for comments on the
choice of reference frame which we have addressed in the Appendix.

\newpage

\begin{appendix}

\section{APPENDIX A. CLOUD EQUILIBRIUM IN THE INERTIAL AND ROTATING FRAMES}

In this appendix we derive the condition of cloud stability (see eq.[10]) in
the galactic reference frame and in the rotating frame moving with the cloud.
We then discuss which equation was used in the paper and the reasons
behind choosing that particular form. 
The coordinate of the cloud with respect to the center of the galaxy is
$\rm{\bf R_{0}}$. The $\rm i^{ith}$ clump in the cloud has a position coordinate
$\rm{\bf R_{i}}$ wrt the galaxy center and $\rm{\bf r_{i}}$ wrt the cloud center,
so that $\rm{\bf R_{i}} = {\bf R_{0}} + {\bf r_{i}}$. The number of clumps in the
cloud is n so that the index i = 1 to n.

\noindent (i) Galactic Frame : The origin of the coordinate system is the center
of the galaxy. For the $\rm i^{ith}$ clump within the cloud the total velocity
$\rm V_{ig}$ is given by,
\be
{\bf V_{ig}} = {\bf V_{0}} + {\bf v_{i}}
\ee
\noindent where $\rm{\bf V_{0}}$ is the rotational velocity of the cloud about the
galaxy center and $\rm {\bf v_{i}}$ is the random motion of the $\rm i^{ith}$ clump.
The kinetic energy of the cloud K, is given by,
\be
2K = M_{c}{V_{0}}^{2} + \sum_{i=1}^{n}m_{i}{v_{i}}^{2}
      + 2{\bf V_{0}.}\left(\sum_{i=1}^{n}{\bf p_{i}}\right)
\ee
\noindent But when the the last term is averaged over time, the random velocities of
the clumps makes the term average out to zero. Hence equation(A2) reduces to,
\be
2\la K\ra =  M_{c}{V_{0}}^{2} + \la\sum_{i=1}^{n}m_{i}{v_{i}}^{2}\ra
\ee
\noindent To determine the right hand side of equation (8), we first determine the
force on the $\rm i^{ith}$ clump, which is the sum of the galactic potential and
the self gravity of the cloud.
\be
{\bf F_{i}} = -\left(\frac{\partial\Phi}{\partial R}\right)_{i}m_{i}{\bf\hat{R_{i}}}
                - \sum_{j=1}^{n}\frac{Gm_{i}m_{j}}{{\mid r_{ij}\mid}^{2}}{\bf r_{ij}}
\ee
\noindent where $\rm {\bf r_{ij}} = ({\bf r_{i}} - {\bf r_{j}})$. 
Expanding the force term $\left(\frac{\partial\Phi}{\partial R}\right)_{i}$
about the
radius $\rm R_{0}$ gives the tidal term so that $\sum_{i=1}^{n} {\bf{F_{i}}.r_{i}}$
reduces to,
\be
\sum_{i=1}^{n} {\bf{F_{i}}.{r_{i}}} =
        -\frac{{V_0}^{2}}{{R_0}^{2}}{\bf\hat{R_{i}}}.\sum_{i=1}^{n}m_{i}{\bf r_{i}}
   + T_{0}\sum_{i=1}^{n}m_{i}{r_{i}}^{2} -
   {{\sum_{i=1}^{n}\sum_{j=1}^{n}}\atop{\scriptstyle i\neq j}}\frac{Gm_{i}m_{j}}{\mid r_{ij}\mid}
\ee
\noindent Using the relation $\rm M_{c}{\bf R_{0}} = \sum_{i=1}^{n}m_{i}{\bf r_{i}}$,
the equation of virial equilibrium for the cloud reduces to,
\be
\la \sum_{i=1}^{n}m_{i}{v_{i}}^{2}\ra = 
   {{\sum_{i=1}^{n}\sum_{j=1}^{n}}\atop{\scriptstyle i\neq j}}\frac{Gm_{i}m_{j}}{\mid r_{ij}\mid}
                               - T_{0}\la\sum_{i=1}^{n}m_{i}{r_{i}}^{2}\ra
\ee
\noindent We have used this form of the virial equation (see eq.[10]) to determine the
minimum cloud density in a compressive tidal field.

\noindent (ii) Rotating frame : In this case the coordinate system is centered in the
cloud so that the $\rm i^{ith}$ clump has a position vector ${\bf r_{i}}$. The force
in the rotating frame is given by,
\be
{\bf F_{i}} = - m_{i}\left(\frac{\partial\Phi}{\partial R}\right)_{i}{\bf\hat{R_{i}}}
         - \sum_{j=1}^{n}\frac{Gm_{i}m_{j}}{{\mid r_{ij}\mid}^{2}}{\bf r_{ij}}
         - 2m_{i}({\bf\omega_{0}}\times{\bf v_{ri}})
         - m_{i}{\bf\omega_{0}}\times({\bf\omega_{0}}\times{\bf R_{i}})
\ee
\noindent where ${\bf\rm\omega_{0}}$ is rotation velocity of the cloud about the galaxy cente
r
i.e. $\rm\mid{\bf\omega_{0}}\mid = \frac{V_{0}}{R_{0}}$ and $\rm{\bf v_{ri}}$
is the random motion of the
$\rm i^{ith}$ clump in the rotating frame. The third term on the right hand side of the
equation is the Coriolis force and the fourth term is the centrifugal force. We expand
the force term $\rm\left(\frac{\partial\Phi}{\partial R}\right)_{i}$ about $\rm R_{0}$ and
put $\rm{\bf R_{i}} = {\bf R_{0}} + {\bf r_{i}}$. Since
$\rm\left(\frac{\partial\Phi}{\partial R}\right)_{0} = {\omega_{0}}^{2}R_{0}$ and
$\rm{\bf\omega_{0}}\times({\bf\omega_{0}}\times{\bf R_{0}}) =
  - {\omega_{0}}^{2}R_{0}{\bf\hat{R_{0}}}$, the force $\rm{\bf F_{i}}$ becomes,
\be
{\bf F_{i}} = T_{0}m_{i}{\bf r_{i}}
       - \sum_{j=1}^{n}\frac{Gm_{i}m_{j}}{{\mid r_{ij}\mid}^{2}}{\bf r_{ij}}
       - 2m_{i}({\bf\omega_{0}}\times{\bf v_{ri}})
       - m_{i}{\bf\omega_{0}}\times({\bf\omega_{0}}\times{\bf r_{i}})
\ee
\noindent Hence, ${\bf F_{i}.r_{i}}$ is given by,
\be
{\bf F_{i}.r_{i}} = T_{0}m_{i}{r_{i}^{2}} -
         - \sum_{j=1}^{n}\frac{Gm_{i}m_{j}}{{\mid r_{ij}\mid}^{2}}{\bf r_{ij}.r_{i}}
         - 2m_{i}({\bf\omega_{0}}\times{\bf v_{ri}}).{\bf r_{i}}
         - m_{i}{\bf\omega_{0}}\times({\bf\omega_{0}}\times{\bf r_{i}}).{\bf r_{i}}
\ee
\noindent Now, $\rm{\bf\omega_{0}}\times({\bf\omega_{0}}\times{\bf r_{i}.}){\bf r_{i}}
= -({\bf\omega_{0}\times r_{i}})^{2}$ and
$\rm({\bf\omega_{0}\times v_{ri}}){\bf .r_{i}} = - {\bf\omega_{0}.}({\bf r_{i}\times v_{ri}})
$.
So equation (A9) becomes,
\be
{\bf F_{i}.r_{i}} = T_{0}m_{i}{r_{i}^{2}}
               - \sum_{j=1}^{n}\frac{Gm_{i}m_{j}}{{\mid r_{ij}\mid}^{2}}{\bf r_{ij}.r_{i}}
                 + 2{\bf\omega_{0}.}(m_{i}{\bf r_{i}\times v_{ri}})
                 + m_{i}({\bf\omega_{0}\times r_{i}})^{2}
\ee
\noindent Also, $\rm{\bf L_{i}} = m_{i}({\bf r_{i}\times v_{i}})$ is the angular
momentum of the clump measured in the inertial or galactic frame and
$\rm{\bf L_{ir}} = m_{i}({\bf r_{i}\times v_{ri}})$ is the cloud angular momentum as measured
in the rotating frame. The two are related through the following relations (e.g. Goldstein, 1980),
\be
{\bf V_{ig}} = {\bf v_{ri}} + {\bf\omega_{0}\times R_{i}}
\ee
\noindent Since $\rm{\bf V_{ig}} = {\bf V_{0}} + {\bf v_{i}}$,
$\rm{\bf R_{i}} = {\bf R_{0}} + {\bf r_{i}}$  and $\rm{\bf V_{0}} = {\bf\omega_{0}\times R_{0
}}$,
we obtain the relation,
\be
{\bf\omega_{0}.L_{i}} = {\bf\omega_{0}.L_{ir}} + m_{i}(\bf{\omega_{0}\times r_{i}})^{2}
\ee
\noindent However, observations indicate that clouds are practically non-rotating, so the
net angular momentum of a cloud  $\rm\sum_{i}{\bf L_{i}}$ as observed in the inertial 
frame is $\rm\sim 0$ (Blitz, 1991).
Applying this to equation (A10) we get the following relation,
\be
\sum_{i}\la{v_{ri}}^{2}\ra = 
 {{\sum_{i=1}^{n}\sum_{j=1}^{n}}\atop{\scriptstyle i\neq j}}\frac{Gm_{i}m_{j}}{\mid r_{ij}\mid}
   - T_{0}\sum_{i=1}^{n}m_{i}{r_{i}}^{2} + \sum_{i=1}^{n}m_{i}({\bf\omega_{0}\times r_{i}})^{2}
\ee
\noindent So the equation of equilibrium is different from that in the inertial or galactic
frame by one term (see eq.[A6]). However, using the relation
$\rm {v_{ri}}^{2} = ({\bf v_{i}} - {\bf\omega_{0}\times r_{i}})^{2}$, we obtain,
\be
\sum_{i}m_{i}{v_{ri}}^{2} = \sum_{i}m_{i}{v_{i}}^{2}
              + \sum_{i}m_{i}({\bf\omega_{0}\times r_{i}})
\ee
\noindent which when substituted in equation (A13) leads to an identical relation for cloud
equilibrium as obtained in the galactic frame and as expected physically.

We have chosen to use the Galactic or inertial frame equation eq.[A6] and not eq.[A13] 
simply because the
velocity is observed from the galactic reference frame and not from within the
cloud (which is the rotating frame). Hence we use the velocity dispersion
$\rm\la\sum_{i}{v_{i}}^{2}\ra$ and not $\rm\la\sum_{i}{v_{ri}}^{2}\ra$.

\end{appendix}

\newpage

\begin{figure}
\caption{Plot of tidal field per unit mass
vs. radius (pc) for a power-law early-type galaxy (NGC 4697) where 
the field is disruptive at all radii, and for a core-profile  galaxy (NGC 4649)
where the field becomes compressive inside a radius of $\sim$130pc.}
\caption{The steady-state number density for
molecular hydrogen gas in bound, virialized clouds vs. radius (pc)  for 
NGC 4472.
Note that the dashed line separates the outer region of disruptive
tidal field from the inner region of compressive tidal field.}
\caption{Plot of tidal field per unit mass
vs. radius (pc) for an ultraluminous galaxy Arp 193.
The field becomes compressive inside a radius of 220 pc. The
sharp transition is an artifact due to the rotation curve model adopted.}
\end{figure}

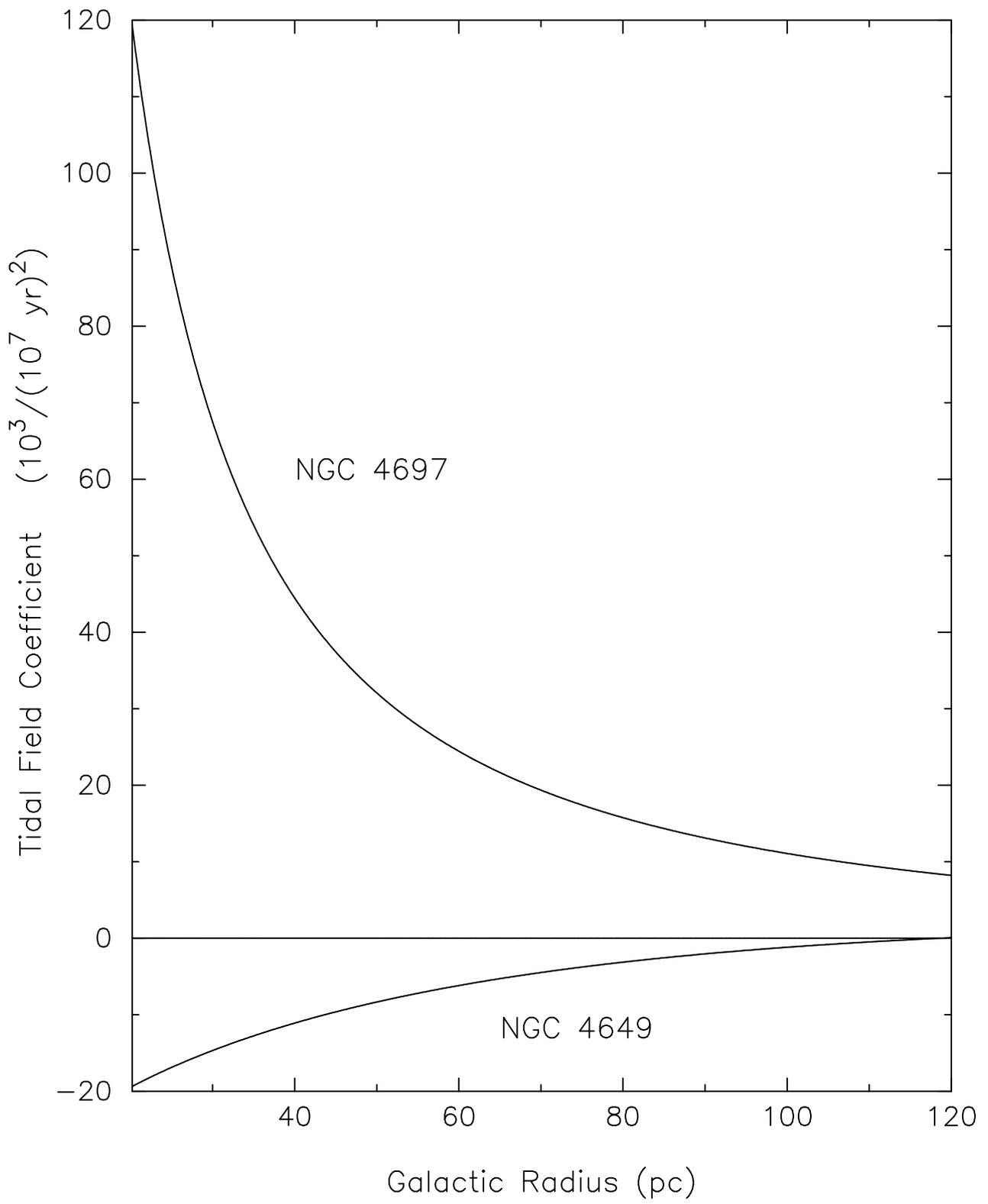
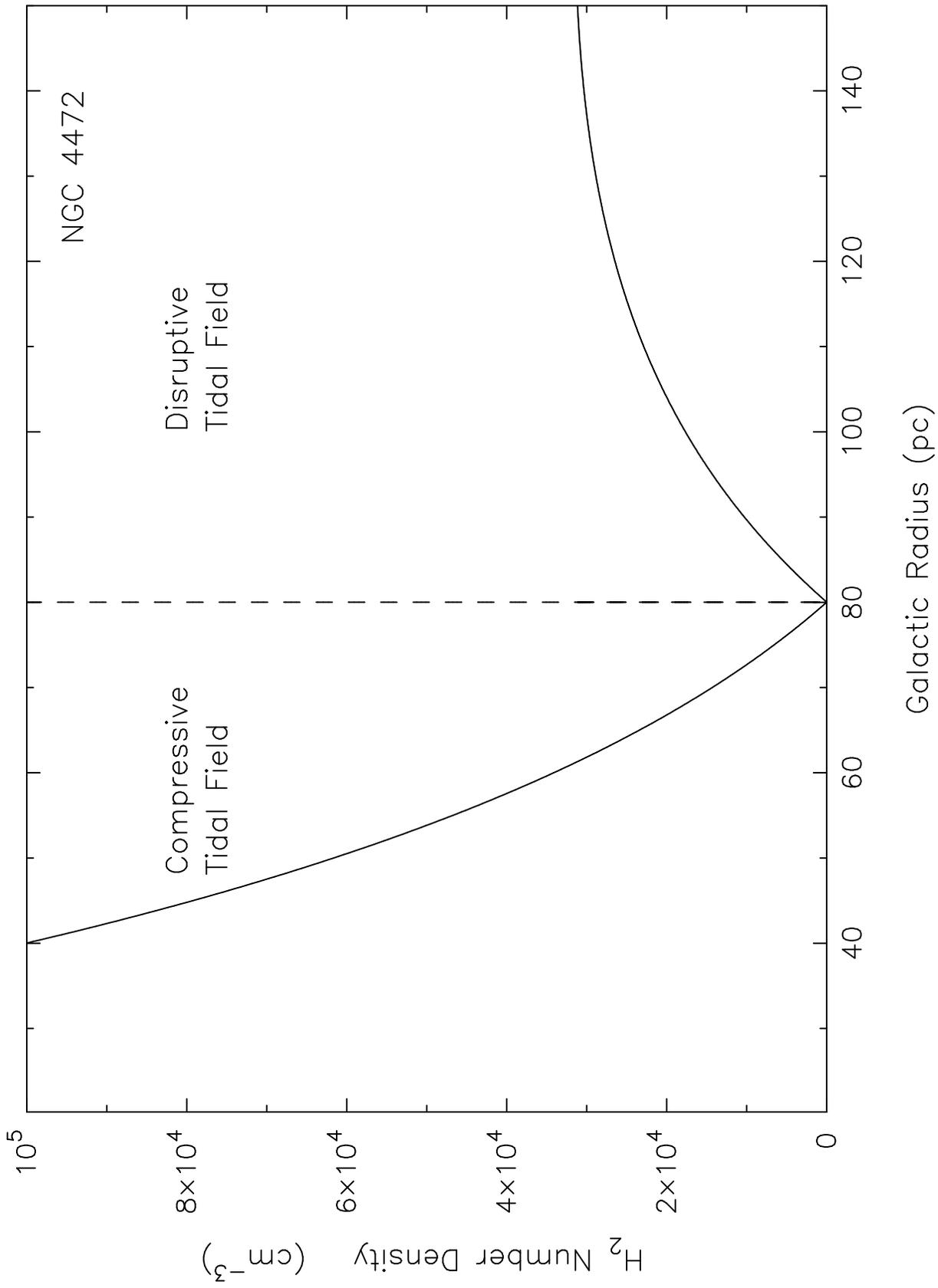
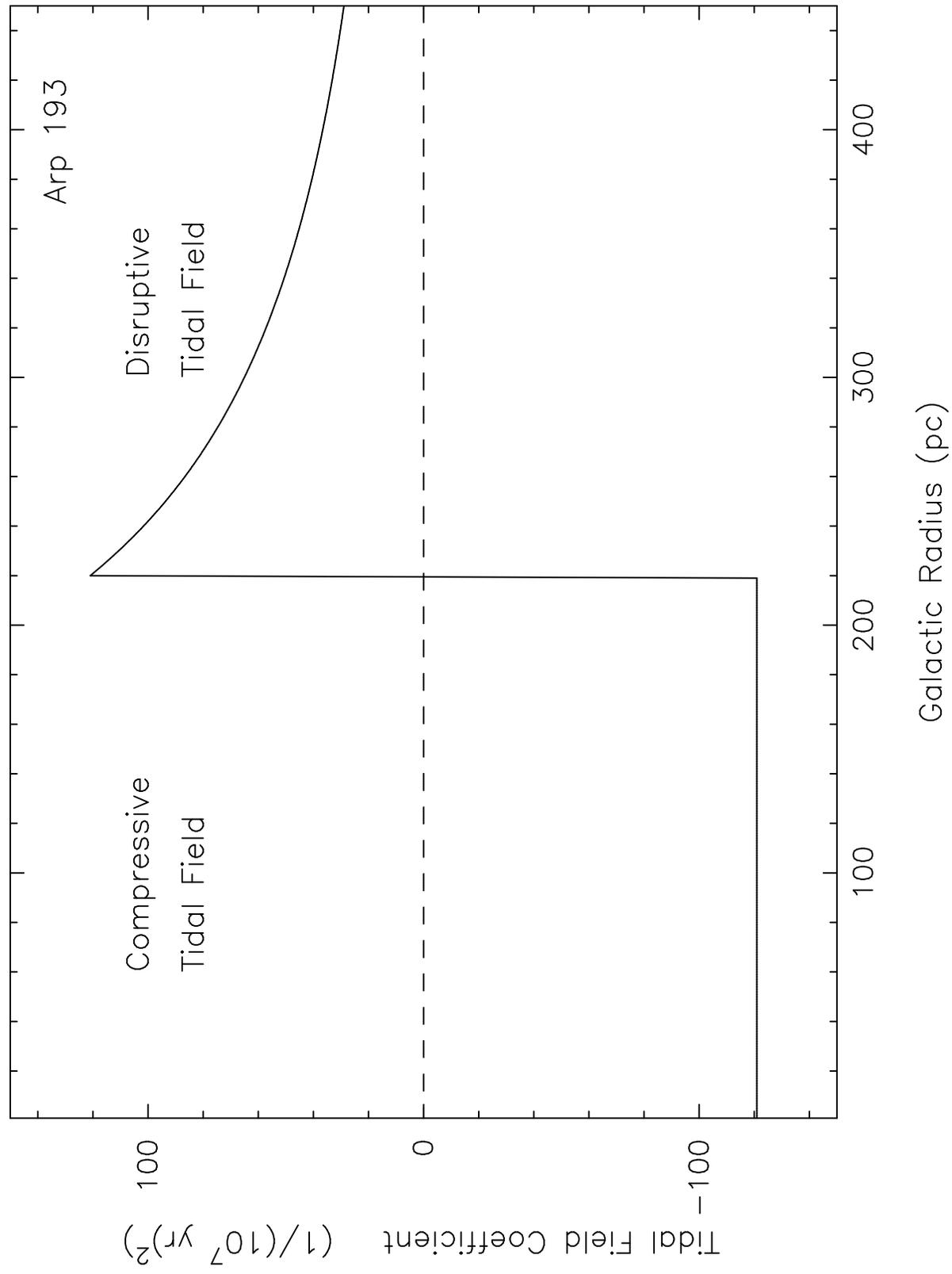
\end{document}